\documentclass[
preprintnumbers,showpacs,amsmath,amssymb,aps,prd]{revtex4}
\usepackage{dcolumn}
\usepackage{amsthm,amscd,mathrsfs}
\usepackage{amsfonts,bm,latexsym}
\def\be{\begin{equation}}
\def\ee{\end{equation}}
\def\bea{\begin{eqnarray}}
\def\eea{\end{eqnarray}}
\def\ba{\begin{array}}
\def\ea{\end{array}}
\def\nn{\nonumber}
\def\p{\partial}

\begin{document}

\title{Thermodynamics and Hawking radiation of five-dimensional rotating charged
G\"{o}del black holes}

\author{Shuang-Qing Wu 
\footnote{Electronic address: sqwu@phy.ccnu.edu.cn}}
\author{Jun-Jin Peng 
\footnote{Electronic address: pengjjph@163.com}}
\affiliation{College of Physical Science and Technology, HuaZhong
Normal University, Wuhan, Hubei 430079, People's Republic of
China}

\begin{abstract}
We study the thermodynamics of G\"{o}del-type rotating charged black holes in
five-dimensional minimal supergravity. These black holes exhibit
some peculiar features such as the presence of closed time-like curves and
the absence of globally spatial-like Cauchy surface. We explicitly compute their
energies, angular momenta, and electric charges that are consistent with the first
law of thermodynamics. Besides, We extend the covariant anomaly
cancellation method, as well as the approach of the effective action, to
derive their Hawking fluxes. Both the methods of the
anomaly cancellation and effective action give the same Hawking fluxes
as those from Planck distribution for blackbody radiation in the background
of the charged rotating G\"{o}del black holes.
Our results further support that Hawking radiation is a quantum phenomenon
arising at the event horizon.
\end{abstract}


\maketitle
\section{Introduction}

G\"{o}del universe is a model that describes the universe with a global rotation.
In four dimensions, the G\"{o}del universe is an exact solution of Einstein field
equation with a negative cosmological constant and homogeneous pressure less
matter, found by G\"{o}del in 1949 \cite{KGodel}. Unlike the usual solutions in
general relativity, this solution possesses some peculiar features such as the
allowance of closed time-like curves and the absence of globally spatial-like
Cauchy surface. It is of great importance for the conceptual development of
general relativity. In recent years, much of interest has been focused on various
generalizations of the four dimensional G\"{o}del universe, particularly in the
context of the five dimensional minimal supergravity theory
\cite{GGHPR,BGHVHarT,GH,CARH,KNG,BDGO,SQWu}. Just as in G\"{o}del's original four
dimensional solution, all the higher dimensional generalized solutions present
closed time-like curves for all times. Furthermore, they can be easily uplifted
to M theory. A remarkable observation also showed that the maximally supersymmetric
analogues of the G\"{o}del universe in \cite{GGHPR} are T-dual to pp-waves
\cite{BGHVHarT}.

Among all the G\"{o}del-type generalizations in five dimensional minimal supergravity,
one solution describing a stationary Kerr black hole embedded in the rotating
G\"{o}del universe was recently found by Gimon and Hashimoto \cite{GH}.
This solution is not required to preserve any supersymmetry, compared with the
supersymmetric one in \cite{GGHPR}. Its various properties
have been intensively investigated
in \cite{TherBC,SteleaSW,KlemmV,KM,QuasiNKA,KerrCFTP,GodelJ}.
Particularly in \cite{TherBC}, Barnich and Comp\`{e}re proposed an effective
method to calculate conserved charges in the G\"{o}del-type background.
They obtained the Kerr G\"{o}del black hole's conserved charges that fulfill
the first law of thermodynamics. The charged generalization of
the Kerr G\"{o}del black hole has been found by one of the authors
\cite{SQWu}. Such a solution is an analytic solution in five dimensional
Einstein field equation coupled with Maxwell and Chern-Simons terms in G\"{o}del
background. We shall refer to it as Einstein-Maxwell-Chern-Simons-G\"{o}del
(EMCS-G\"{o}del) black hole. After getting this black hole solution, it is
very necessary to study its thermodynamical properties. In this paper, we
explicitly compute the mass, angular momenta and electric charge of
the EMCS-G\"{o}del black hole along the lines of \cite{TherBC}. These conserved
charges satisfy the differential first law and the generalized integral Smarr
formula of black hole thermodynamics. However, unlike work \cite{TherBC},
to close the integral Smarr formula, the G\"{o}del parameter
is seen as a thermodynamical variable \cite{SQWu}. For the extremal
EMCS-G\"{o}del black holes, their microscopic entropies can be derived through
Kerr/CFT correspondence \cite{KerrCFTP}.

In the above, we have mentioned that the EMCS-G\"{o}del black hole exhibits
thermodynamical characters. Thus there must exist Hawking radiation at its
event horizon. This quantum phenomenon is actually very universal and can be found
in any geometry background with event horizons. It is regarded as  a clue for seeking
the theory of quantum gravity. Although Hawking radiation
has not yet been observed on laboratory, it has been verified by several
different approaches since Hawking discovered this effect more than thirty
years ago. Recently, Wilczek and his collaborators proposed a new derivation
of Hawking radiation from four dimensional black holes via gravitational and
gauge anomalies \cite{RWanoma,IUW,RotatingAnom}. In their works, Hawking radiation
is treated as a compensating flux to cancel gravitational and gauge anomalies at
the horizon, which arise since the effective field theory becomes two dimensional
and chiral after performing a procedure of dimensional reduction near the horizon
of a black hole. This anomaly cancellation method supports that Hawking radiation
is a common property of the horizon. It is very universal and has been successfully
applied to black objects in various dimensions
\cite{MurataJ,KUPhD,GodelJ,RotAnomB,BonoraC,Sphano,Wueff,BRing,BKulk,Baner,CAstr,GangKul}.
Noticing that the anomalous energy momentum tensors and currents
encompass two types of forms in the two dimensional chiral effective theory,
apart from the consistent form in \cite{RWanoma,IUW,RotatingAnom},
the other type is the covariant one. In \cite{BKulk},
it was argued that Hawking fluxes of energy momentum tensors and gauge currents can
be obtained by cancelling the covariant gravitational anomaly and gauge
anomaly at the horizon. Such an argument makes the original
anomaly cancellation method \cite{RWanoma,IUW,RotatingAnom} more economical
and conceptually cleaner. Based on development in \cite{BKulk},
several extensions can be found in \cite{Baner,CAstr,GangKul}. Especially in \cite{CAstr},
Hawking radiation of black strings in four and higher dimensions has been
studied via covariant anomalies.

A notable feature of the anomaly cancellation method is that the boundary
conditions at the event horizon play an important role in determining the
Hawking fluxes. Indeed, in \cite{BanerKeff}, by only imposing the boundary
condition that the covariant energy momentum tensor and the covariant gauge
current vanish at the horizon, the chiral effective action, which describes
the two dimensional chiral theory near the horizon, has been used to compute
the Hawking fluxes of charged spherically symmetric black holes. This effective
action method is very universal and holds true for other black holes
\cite{GodelJ,Wueff,effappli}. In addition to the chiral effective
action, the normal effective action that induces anomaly free energy momentum
tensors and gauge currents has also reproduced the Hawking fluxes of the
Reissner-Nordstr\"{o}m black hole \cite{RotatingAnom}. A lot of works on
applying the effective action to study Hawking effect can be found in
\cite{LeoRTuna,AShirasakaT,TwoDefac,TwoDefBF}.

In this paper, we investigate the thermodynamics of the EMCS-G\"{o}del
black hole and then generalize the covariant anomaly cancellation method,
as well as the effective action approach, to study its Hawking radiation.
Both the methods present the same Hawking fluxes.
Our results support that Hawking radiation is a universal quantum phenomenon
arising at the event horizon. The remainder of this paper goes as follows.
In section \ref{hfthermo}, we calculate the mass, the angular momenta
and the electric charge of the EMCS-G\"{o}del black hole, which satisfy
the first law of thermodynamics.
In section \ref{hfcova}, we compute the Hawking fluxes by treating them
as compensating fluxes to cancel the covariant gravitational and gauge
anomalies near the horizon. In section \ref{hfeffa}, we reproduce the
Hawking fluxes of the EMCS-G\"{o}del black hole via the approach of
the effective action, including the normal effective action in subsection
\ref{hfeffa1} and the chiral effective action in subsection \ref{hfeffa2}.
The last section is our conclusions.

\section{Thermodynamics of the EMCS-G\"{o}del black hole}\label{hfthermo}

In this section, we study the thermodynamics of the EMCS-G\"{o}del black hole
\cite{SQWu}. Although the main results were presented in \cite{SQWu}, here we
give the explicit calculations by adopting the gauge field whose electric-static
potential vanishes at infinity. Our starting point is the EMCS-G\"{o}del black hole,
which is a non-extremal charged rotating G\"{o}del-type black hole solution
in five-dimensional ungauged minimal supergravity. The relevant Einstein-Maxwell
Lagrangian with Chern-Simons term reads
\be
L=\frac{\sqrt{-g}}{16\pi}(R-F_{\mu\nu}F^{\mu\nu})-\frac{1}{24\pi\sqrt{3}}
   \epsilon^{\lambda\rho\sigma\mu\nu}A_\lambda F_{\rho\sigma}F_{\mu\nu}
 \, , \label{FiveDLagran}
\ee
where $\epsilon^{\lambda\rho\sigma\mu\nu}$ is the five-dimensional tensor density
with $\epsilon^{01234}=-1$, and $F_{\mu\nu}=\partial_\mu A_\nu-\partial_\nu A_\mu$
denotes the abelian field-strength tensor. The Einstein and gauge field equations
of motion derived from Lagrangian (\ref{FiveDLagran}) are
\bea
&&R_{\mu\nu}-\frac{1}{2}g_{\mu\nu}R =2\left(F_{\mu\alpha}F_{\nu}^{~\alpha}
-\frac{1}{4}g_{\mu\nu}F_{\rho\sigma}F^{\rho\sigma} \right) \, , \nn \\
&&\nabla_\nu \left(F^{\mu\nu}+\frac{1}{\sqrt{3}\sqrt{-g}}
\epsilon^{\mu\nu\lambda\rho\sigma}A_\lambda F_{\rho\sigma}\right)
=0 \, . \label{FGMotionEq}
\eea

Parameterized by four constants $(\mu, a, q, j)$, which correspond to the mass,
the angular momentum, the electric charge and the scale of the G\"{o}del background,
respectively, the EMCS-G\"{o}del black hole satisfying Eq. (\ref{FGMotionEq})
takes the form \cite{SQWu}
\bea
 ds^2 &=& -f(r)\Big [dt+\frac{h(r)}{f(r)}(d\phi+\cos \theta d\psi)\Big]^2
  +\frac{1}{4}r^2(d\theta^2+\sin^2 \theta d\psi^2) \nn \\
 && +\frac{dr^2}{V(r)} +\frac{r^2V(r)}{4f(r)}(d\phi+\cos \theta d\psi)^2 \, ,
 \label{GBHansatz} \\
 A &=& B(r)dt +C(r)(d\phi+\cos \theta d\psi) \, , \label{Gaugeanz}
\eea
where
\bea
 f(r) &=& 1 -\frac{2\mu}{r^2} +\frac{q^2}{r^4} \, , \nn \\
 h(r) &=& jr^2 +3jq +\frac{(2\mu -q)a}{2r^2} -\frac{q^2a}{2r^4} \, , \nn \\
 V(r) &=& 1 -\frac{2\mu}{r^2} +\frac{8j(\mu +q)\big[a +2j(\mu +2q)\big]}{r^2}
      +\frac{2(\mu -q)a^2}{r^4}\nn \\
 && +\frac{q^2\big[1 -16ja -8j^2(\mu +3q)\big]}{r^4} \, , \nn \quad \\
 B(r) &=& \frac{\sqrt{3}q}{2r^2} \, , \quad
 C(r) = \frac{\sqrt{3}}{2}\Big(jr^2 +2jq -\frac{qa}{2r^2}\Big) \, . \label{fhvBC}
\eea
In the above equations, the Euler angles $\theta$, $\psi$ and $\phi$ run over the ranges
$0$ to $\pi$, $0$ to $2\pi$ and $0$ to $4\pi$, respectively. The line element
(\ref{GBHansatz}) is the charged generalization of the Kerr G\"{o}del black hole.
It is asymptotically rotating. Just as its uncharged counterpart, it
exhibits the peculiar features such as the presence of closed time-like curves and
the absence of globally spatial-like Cauchy surface.
When the electric charge parameter $q=0$, it returns to the Kerr G\"{o}del black
hole in \cite{GH}, whose Hawking radiation has been investigated via the covariant
anomalies and effective action \cite{GodelJ}. The angular velocities and the
electro-static potential of the EMCS-G\"{o}del black hole are given by
\bea
\Omega(r) &=&\Omega_\phi = h(r)/U(r), \qquad \qquad \Omega_\psi = 0 \, , \\
\Phi &=& \ell^\mu A_\mu = B(r) + \Omega_\phi C(r) \, , \label{elepote}
\eea
where
\bea
U(r) &=& \frac{r^2V(r)-4h^2(r)}{4f(r)} \nn \\
&=& -j^2r^2(r^2 +2\mu +6q) +3jqa
 +\frac{(\mu -q)a^2}{2r^2} -\frac{q^2a^2}{4r^4} +\frac{r^2}{4} \, ,
\eea
and the corotating vector $\ell =\partial_t +\Omega(r) \partial_\phi$. With
help of this vector, the surface gravity $\kappa$ is defined by
$\kappa^2=-\frac{1}{2}\ell_{\mu;\nu}\ell^{\mu;\nu}|_{r=r_+}$, where the outside
event horizon $r_+$ is determined by equation $V(r_+)=0$ and reads
\bea
r^2_+ &=& \mu -4j(\mu +q)a -8j^2(\mu +q)(\mu +2q) +\sqrt{\delta} \, , \nn \\
\delta &=& [\mu -q -8j^2(\mu +q)^2] \nn \\
&& \times[\mu +q -2a^2 -8j(\mu +2q)a -8j^2(\mu +2q)^2] \, .
\eea
Hence Hawking temperature via the surface gravity formula is read off as
\be
T_H =\frac{\kappa}{2\pi}
    =\frac{r_+V'(r_+)}{8\pi\sqrt{U(r_+)}} \, . \label{HTsurgrav}
\ee
Here, and in what follows, the prime $'$ denotes the derivative with respect to
the radial coordinate $r$. The entropies via the Bekenstein-Hawking area law are
\be
S = \pi^2r^2_+\sqrt{U(r_+)}  \, . \label{BHentro}
\ee
It is worth noting that the electro-static potential (\ref{elepote}) is not zero
at infinity, but $\Phi_\infty =-\sqrt{3}/2$, since the G\"{o}del universe possesses
a global rotation at infinity. In order to make the electro-static potential
vanish at infinity, we can rescale the gauge field (\ref{Gaugeanz}) as
\be
A= \hat{B}(r)dt +C(r)(d\phi+\cos \theta d\psi) \, , \label{ReGauge}
\ee
where $\hat{B}(r) =B(r)+\sqrt{3}/2$. We shall adopt Eq. (\ref{ReGauge}) for all the
calculations related to gauge fields.

Now, we compute the mass, angular momenta, and electric charge of the
EMCS-G\"{o}del black hole. Because of the presence of closed time-like
curves and the special asymptotical structure
of the G\"{o}del-type black hole, naive application of the traditional
approaches, such as the methods of Komar integral, the usual Abbott-Deser
construction and the covariant phase
space \cite{IyerWald}, fails to give conserved charges in agreement with the first
law of thermodynamics. In \cite{TherBC}, a new method, based on cohomological
techniques \cite{BBrandt}, has been successfully used to derive the
conserved charges of the Kerr G\"{o}del black hole. This method is also
applicable to the
EMCS-G\"{o}del black hole. Our computation follows work \cite{TherBC}.
Here we only give the formulas closely relevant to our calculations.
For more details see \cite{TherBC,BBrandt}.

Let $\varphi^i=(g_{\mu\nu},A_\mu)$ denote the fields of the five-dimensional
ungauged minimal supergravity. $\bar{\varphi}^i=(\bar{g}_{\mu\nu},\bar{A}_\mu)$
is any fixed reference solution of the motion equtions in Eq. (\ref{FGMotionEq}).
Consider the linearized theory for the variables
$\delta\varphi^i=\varphi^i-\bar{\varphi}^i
=(\delta g_{\mu\nu},\delta A_\mu)=(h_{\mu\nu},a_\mu)$. The equivalence classes
of conserved 3-forms of this linearized theory are in correspondence with
equivalence classes of field dependent gauge parameters $\xi^\mu (x)$ and
$\Lambda (x)$ satisfying the reducibility equations \cite{TherBC}
\bea
&&\mathcal{L}_\xi \bar{g}_{\mu\nu} =0 \, , \nn \\
&&\mathcal{L}_\xi \bar{A}_\mu +\partial_\mu \Lambda =0 \, . \label{ReduciEq}
\eea
Each pair of solutions $(\xi,\Lambda)$ of Eq. (\ref{ReduciEq}) is associated
with a conserved 3-form $k_{\xi,\Lambda}[\delta\varphi,\bar{\varphi}]$
that can be obtained by computing the weakly vanishing Noether currents
related to the gauge transformations. Obviously, when $\xi$ is a Killing
vector $\bar{\xi}$ of the background $\bar{\varphi}$ and $\Lambda$ is a
constant $c$, Eq. (\ref{ReduciEq}) holds. For the solutions $(\bar{\xi},0)$,
the conserved 3-form $k_{\xi,\Lambda}$ can be decomposed as
$k_{\bar{\xi},0}=k_{\bar{\xi}}^{gr}+k_{\bar{\xi}}^{em}+k_{\bar{\xi}}^{CS}$,
where $k_{\bar{\xi}}^{gr}$, $k_{\bar{\xi}}^{em}$ and $k_{\bar{\xi}}^{CS}$
are the contributions from gravitation, electromagnetism and the Chern-Simons
term, respectively. $k_{\bar{\xi}}^{gr}$ is defined by
\be
k_{\bar{\xi}}^{gr}[h,\bar{g}]=-\delta K_{\bar{\xi}}^K
-\bar{\xi}\cdot \Theta^{gr} \, ,
\ee
where the Komar 3-form
\bea
K_{\bar{\xi}}^K &=&\frac{\sqrt{-g}}{192\pi}(\nabla^\mu\bar{\xi}^\nu-\nabla^\nu\bar{\xi}^\mu)
\epsilon_{\mu\nu\lambda\rho\sigma} dx^\lambda\wedge dx^\rho \wedge dx^\sigma
\, , \\
\Theta^{gr} &=&\frac{\sqrt{-\bar{g}}}{384\pi}(\bar{\nabla}_\alpha h^{\mu\alpha}
-\bar{\nabla}^\mu h)
\epsilon_{\mu\nu\lambda\rho\sigma}dx^\nu \wedge dx^\lambda\wedge dx^\rho \wedge dx^\sigma
\, ,
\eea
and $\bar{\xi}\cdot = \bar{\xi}^\mu \frac{\partial}{\partial (dx^\mu)}$. The electromagnetic
contribution $k_{\bar{\xi}}^{em}$ is similar with $k_{\bar{\xi}}^{gr}$, which reads
\be
k_{\bar{\xi}}^{em}[a,h;\bar{A},\bar{g}]=-\delta Q_{\bar{\xi},0}^{em}
-\bar{\xi}\cdot \Theta^{em} \, ,
\ee
where
\bea
Q_{\bar{\xi},c}^{em} &=&\frac{\sqrt{-g}}{48\pi}
[(\bar{\xi}^\alpha A_\alpha+c)F^{\mu\nu}]
\epsilon_{\mu\nu\lambda\rho\sigma} dx^\lambda\wedge dx^\rho \wedge dx^\sigma
\, , \\
\Theta^{em} &=&\frac{\sqrt{-\bar{g}}}{96\pi}(\bar{F}^{\alpha\mu}
a_\alpha)
\epsilon_{\mu\nu\lambda\rho\sigma}dx^\nu \wedge dx^\lambda\wedge dx^\rho \wedge dx^\sigma
\, .
\eea
The contribution from the Chern-Simons term is
\be
k_{\bar{\xi}}^{CS}[a,\bar{A}] = \frac{1}{2\sqrt{3}\pi}
(\bar{\xi}^\alpha \bar{A}_\alpha)a_\rho\bar{F}_{\mu\nu}
dx^\rho\wedge dx^\mu \wedge dx^\nu
\, .
\ee
For the solution $(0,1)$, which corresponds to the contribution from the
electric charge, the conserved 3-form
\be
k_{0,1}[a,h;\bar{A},\bar{g}] = -\delta (Q_{0,1}^{em}+\mathbb{J}) \, ,
\ee
where
\be
\mathbb{J}=-\frac{1}{4\sqrt{3}\pi}
A_\rho F_{\mu\nu}
dx^\rho\wedge dx^\mu \wedge dx^\nu
\, .
\ee

Take into account a path $\gamma$ in the space of solutions that
interpolates between a given solution $\varphi$ and the background
solution $\bar{\varphi}$. Let $d_V\varphi$ be a one-form in the field
space. As long as the pair
$(\bar{\xi},c)$ satisfy Eq. (\ref{ReduciEq}) for all solutions along
this path, we can get a closed 3-form
\be
K_{\bar{\xi},c} =\int_\gamma k_{\bar{\xi},c}[d_V\varphi;\varphi] \, ,
\label{closedK}
\ee
i.e. $dK_{\bar{\xi},c}=0$ in a four-dimensional hypersurface $\Sigma$.
Using Eq. (\ref{closedK}), one can define conserved charges
\be
Q_{\bar{\xi},c} = \oint_S K_{\bar{\xi},c}, \label{defcharge}
\ee
where the three-dimensional closed surface $S$ is the boundary of the
hypersurface $\Sigma$.

Next we turn our attention to calculate the conserved charges of
the EMCS-G\"{o}del black hole via (\ref{defcharge}). The mass
is computed as
\bea
M &=& \oint_S K_{\partial/\partial_t,0} \nn \\
 &=&\frac{3}{4}\pi(m+q)-4\pi(m+q)ja-8\pi(m+2q)(m+q)j^2 \, .
\eea
The angular momentum along the $\phi$ direction
\bea
J_\phi &=&-\oint_S K_{\partial/\partial_\phi,0} \nn \\
 &=&\frac{1}{2}\pi\Big\{a\Big[m-\frac{q}{2}-2(m-q)aj-8(m^2+mq-2q^2)j^2\Big] \nn \\
 &&-3jq^2+8(3m+5q)j^2q^2\Big\} \, ,
\eea
while the one with respect to the coordinate $\psi$ is zero. The electric
charge is given by
\bea
Q &=& \oint_S K_{0,1} \nn \\
 &=&\frac{\sqrt{3}}{2}\pi[q-4(m+q)aj-8(m+q)qj^2] \, .
\eea
The electric charge can also be computed through
\be
Q=\frac{1}{4 \pi}\int_{S_3} \left(\frac{1}{12}\sqrt{-g}F^{\alpha\beta}
\epsilon_{\alpha\beta\rho\mu\nu}-\frac{1}{\sqrt{3}}
A_\rho F_{\mu\nu} \right)
dx^\rho\wedge dx^\mu \wedge dx^\nu
\, ,
\ee
where the integration is performed on the 3-sphere at infinity.
All the conserved charges are consistent with the first law of thermodynamics
\bea
dM &=& T_HdS + \Omega_+ dJ_\phi +\Phi_+ dQ +W dj \, , \\
\frac{2}{3}M &=& T_H S + \Omega_+ J_\phi + \frac{2}{3}\Phi_+ Q -\frac{1}{3}W j \, ,
\eea
where $\Omega_+=\Omega(r_+)$ and $\Phi_+ =\hat{B}(r_+) +\Omega(r_+)C(r_+)$ are
the angular velocity and the electro-static
potential at the event horizon, and
\be
W=2\pi(m+q)[a+2j(m+2q)]  \nn
\ee
is the generalized force conjugate to the G\"{o}del parameter $j$ since we
have considered $j$ as a thermodynamical variable to close the expression
of the integral Bekenstein-Smarr formula.

\section{Hawking fluxes and covariant anomalies}\label{hfcova}

In this section, we shall investigate Hawking radiation of the EMCS-G\"{o}del
black hole \cite{SQWu} via the covariant gravitational and gauge anomaly
cancellation method \cite{BKulk} developed on basis of \cite{RWanoma,IUW,RotatingAnom}.
The same results will be obtained if we adopt the consistent anomaly cancellation
method in \cite{RWanoma,IUW,RotatingAnom}. Before our proceeding, it is
necessary for us to briefly review this approach. By performing the technique of
dimensional reduction, the massless scalar field near the horizon can be effectively
described by a collection of scalar fields in the background of (1+1)-dimensional
spacetime. Thereby we can treat the higher dimensional theory as a (1+1)-dimensional
effective theory near the horizon. If we omit the classically irrelevant ingoing modes
inside the horizon, the two dimensional effective theory becomes chiral. Such a
chiral theory exhibits covariant gravitational and gauge anomalies. Imposing
the boundary condition that the covariant energy momentum tensor and current vanish
at the horizon, we can get fluxes that just cancel these anomalies and are
identified with Hawking fluxes for the energy momentum tensor and charges.

We first implement a process of dimensional reduction by considering the free part
of the action for a scalar massless complex field in the background of metric
(\ref{GBHansatz}) and gauge field (\ref{ReGauge}). We have
\bea
 S[\varphi] &=& \frac{1}{2}\int d^5x\varphi^*\mathcal{D}_{\mu}\big(\sqrt{-g}
  g^{\mu\nu} \mathcal{D}_{\nu}\varphi\big) \nn \\
 &=& \frac{1}{16}\int dtdrd\theta d\phi d\psi\sin\theta\varphi^*\Big\{-\frac{4rU(r)}{V(r)}
  \big(\mathcal{D}_t +\Omega(r)\mathcal{D}_\phi\big)^2
  +\p_r \big[r^3V(r)\p_r\big]  \nn \\
 &&\quad +\frac{r^3}{U(r)}\mathcal{D}_\phi^2
  +4r\Big[\frac{1}{\sin\theta}\p_\theta (\sin\theta\p_\theta)
  +\frac{(\partial_\psi -\cos\theta\partial_\phi)^2}{\sin^2\theta}\Big]
   \Big\}\varphi
  \, , \label{Saction}
\eea
where $\mathcal{D}_{\mu} = \p_\mu +ieA_\mu$. After performing a partial wave
decomposition $\varphi =\sum_{lmn}\varphi_{lmn}(t, r)\exp(im\phi +in\psi)
\Theta_{lmn}(\theta)$, where the spin-weighted spheroidal functions
$\Theta_{lmn}(\theta)$ satisfy
\be
\Big[\frac{1}{\sin\theta}\p_\theta (\sin\theta\p_\theta)
  -\frac{(n -m\cos\theta)^2}{\sin^2\theta}
  +l(l+1)-m^2\Big]\Theta_{lmn}(\theta) =0 \, ,
\ee
and only keeping the dominant terms near the horizon, the
action (\ref{Saction}) becomes
\bea
 S[\varphi] &\simeq& \frac{1}{8}\sum_{lmn}
 \int dtdr ~r^2\sqrt{U(r)}\varphi^*_{lmn}
  \Big\{-\frac{1}{F(r)}\big[\p_t +ie \big(\hat{B}(r)+\Omega(r) C(r)\big) \nn \\
&&\quad +im\Omega(r)\big]^2
  +\p_r \left[F(r)\p_r\right]\Big\}\varphi_{lmn} \, . \label{Saction2}
\eea
In Eq. (\ref{Saction2}), we have defined $2F(r)=rV(r)U(r)^{-1/2}$.
Thereby the physics near the horizon can be described by an infinite set of
effective massless fields on a ($1 + 1$)-dimensional spacetime
with the metric and the gauge potential
\bea
 ds^2 &=& -F(r)dt^2 +\frac{dr^2}{F(r)} \, ,  \label{effmetr} \\
 \mathcal{A}_t &=& e\mathcal{A}_t^{(0)} +m\mathcal{A}_t^{(1)}
 = e \big[\hat{B}(r)+\Omega(r) C(r)\big] +m\Omega(r) \, , \quad
 \mathcal{A}_r =0\, ,  \label{effgau}
\eea
where $\mathcal{A}_t(\infty)=0$, and $F( \infty)= F^\prime(\infty)
=F^{\prime\prime}(\infty)=0$.
In such a two dimensional effective theory, the $t$-component of the gauge
field $\mathcal{A}$ contains two types of U(1) fields. the gauge field
$\mathcal{A}_t^{(0)}$ comes from the original electric field (\ref{ReGauge}), while
$\mathcal{A}_t^{(1)}$ can be interpreted as an induced $U(1)$ gauge field
from the axial isometry in the $\phi$ direction. The azimuthal quantum number $m$
for each partial wave serves as charges of the gauge field $\mathcal{A}_t^{(1)}$.

Next, we pay our attention to derive the currents of the gauge field
(\ref{effgau}) via covariant gauge anomaly. In our case, there are two $U(1)$
gauge symmetries yielding two gauge currents $J^{(0)r}$ and $J^{(1)r}$,
corresponding to the gauge potentials $\mathcal{A}_t^{(0)}$ and
$\mathcal{A}_t^{(1)}$, respectively. Except for different types of charges,
both the gauge potentials are essentially consistent with each other. Thus
we only give an explicit derivation of the current $J^{(0)r}$. $J^{(1)r}$
can be obtained by a similar procedure.

Due to the anomaly cancellation method, the gauge current behaves differently
in the range outside the horizon and that near the horizon. In the former, namely,
the range $r\in[r_+ +\varepsilon, +\infty)$, the current $J_{(O)}^{(0)\mu}$ is
anomaly free and takes the conserved form
\be
\nabla_\mu J_{(O)}^{(0)\mu} = 0 \, , \label{consgauc}
\ee
while in the range near the horizon $(r\in[r_+, r_+ +\varepsilon])$,
because of the breakdown of the classical gauge symmetry, the current $J_{(H)}^{(0)\mu}$
satisfies the anomaly Ward identity \cite{IUW,RotatingAnom,BKulk}
\be
 \nabla_\mu \frac{1}{e}J_{(H)}^{(0)\mu}
  = \frac{-1}{4\pi\sqrt{-g}} \epsilon^{\alpha\beta}\mathcal{F}_{\alpha\beta} \, ,
 \label{anougauc}
\ee
where $\epsilon^{\alpha\beta}$ is an antisymmetry tensor density with
$\epsilon^{tr} = -\epsilon_{tr} = 1$ and $\mathcal{F}_{\alpha\beta}
= \p_\alpha\mathcal{A}_\beta -\p_\beta\mathcal{A}_\alpha$. Solving Eqs.
(\ref{consgauc}) and (\ref{anougauc}), we have
\bea
 \sqrt{-g}J_{(O)}^{(0)r} &=& c_O^{(0)} \, , \nn  \\
 \sqrt{-g}J_{(H)}^{(0)r} &=& c_H^{(0)} +\frac{e}{2\pi}
  \big[\mathcal{A}_t(r) -\mathcal{A}_t(r_+)\big] \, ,
\eea
where the charge flux $c_O^{(0)}$ and $c_H^{(0)}$ are two integration constants,
which denote the current at infinity and the one at the horizon,
respectively. Introducing two step functions $\Theta(r) =
\Theta(r -r_+ -\varepsilon)$ and
$H(r) = 1 -\Theta(r)$ to write the total current as
\be
 J^{(0)\mu} = J_{(O)}^{(0)\mu}\Theta(r) +J_{(H)}^{(0)\mu}H(r) \, ,
 \label{totagf}
\ee
we find that the Ward identity becomes
\be
 \p_r\big[\sqrt{-g}J^{(0)r}\big] = \p_r\big(\frac{e}{2\pi}\mathcal{A}_tH\big)
 +\big\{\sqrt{-g}\big[J_{(O)}^{(0)r} -J_{(H)}^{(0)r}\big]
 +\frac{e}{2\pi}\mathcal{A}_t\big\}\delta(r -r_+ -\varepsilon) \, .
\ee
In order to make the current preserve the gauge symmetry, the first term
in the above equation must be cancelled by the classically irrelevant
ingoing modes while the second term should vanish at the horizon, which
yields
\be
 c_O^{(0)} = c_H^{(0)} -\frac{e}{2\pi} \mathcal{A}_t(r_+) \, , \qquad
 \mathcal{A}_t(r_+) = e\big[\hat{B}(r_+)+\Omega(r_+) C(r_+)\big] +m\Omega(r_+)\, .
\ee
Further imposing the boundary condition that the covariant current
vanishes at the horizon, namely, $c_H^{(0)} = 0$, then the charge
flux corresponding to the gauge potential $\mathcal{A}_t^{(0)}$ is
given by
\be
c_O^{(0)} = -\frac{e}{2\pi}\mathcal{A}_t(r_+) \, .
\ee
Following the analysis of computing $c_O^{(0)}$ step by step, the
current with respect to the gauge potential $\mathcal{A}_t^{(1)}$
reads
\be
 c_O^{(1)} = -\frac{m}{2\pi}\mathcal{A}_t(r_+) \, .
\ee
From Eq. (\ref{anougauc}), one can see that $J^{(0)r}$ and $J^{(1)r}$ are not
independent for each oter but there exists the relation $\frac{1}{e}J^{(0)r}
= \frac{1}{m}J^{(1)r} = \mathcal{J}^{r}$ between them,
where $\mathcal{J}^{\mu}$ satisfies the covariant gauge anomaly
equation
\be
 \nabla_\mu \mathcal{J}_{(H)}^{\mu}
 = \frac{-1}{4\pi\sqrt{-g}}\epsilon^{\alpha \beta}
 \mathcal{F}_{\alpha \beta} \, ,
\ee
near the horizon. By analogy, the current out of the horizon can be solved as
\be
c_O = -\frac{1}{2\pi}\mathcal{A}_t(r_+) \, . \label{gcurrca}
\ee

With the expression of the charge flux in hand, we now consider the energy
momentum flux in the way similar to the gauge anomaly. Near the horizon,
if we eliminate the quantum effect of the ingoing modes, the invariance
under general coordinate transformation will break down. Thus the two
dimensional effective field theory will exhibit a gravitational anomaly.
For the right-handed fields, the covariant gravitational anomaly has the
form \cite{BKulk}
\be
 \nabla_{\mu}T_{~\nu}^{\mu}
 = \frac{1}{96\pi\sqrt{-g}}\epsilon_{\nu\mu}\p^{\mu}R
 = \frac{1}{\sqrt{-g}}\p_{\mu}N_{~\nu}^{\mu} \, .
\ee
In the case of a background spacetime with the effective metric (\ref{effmetr}),
the anomaly is timelike ($\nabla_{\mu}T_{~t}^{\mu}=0$), and
\be
 N_{~t}^{r} = \frac{1}{192\pi}\big(2FF^{\prime\prime}
  -F^{\prime 2}\big) \, .
\ee
Because of the presence of the external gauge field $\mathcal{A}$, the energy
momentum tensor outside the horizon does not take the conserved form but
satisfies the Lorentz force law
\be
 \nabla_{\mu}T_{(O)\nu}^{\mu} = \mathcal{F}_{\mu\nu}\mathcal{J}_{(O)}^{\mu} \, ,
 \label{gat1}
\ee
while the energy momentum near the horizon obeys the anomalous Ward identity
after adding the gravitational anomaly,
\be
 \nabla_{\mu}T_{(H)\nu}^{\mu} = \mathcal{F}_{\mu\nu}\mathcal{J}_{(H)}^{\mu}
 +\frac{1}{96\pi\sqrt{-g}}\epsilon_{\nu\mu}\p^{\mu}R \, . \label{gat2}
\ee
Solving both the equations (\ref{gat1}) and (\ref{gat2}) for the
$\nu = t$ component, we get
\begin{subequations}
\begin{align}
 \sqrt{-g}T_{(O)t}^{r} &= a_O +c_O\mathcal{A}_t(r) \, , \label{aoah1} \\
 \sqrt{-g}T_{(H)t}^{r} &= a_H +\Big[c_O\mathcal{A}_t(r)
 +\frac{1}{4\pi}\mathcal{A}_t^2(r)
 +N_{~t}^{r}\Big]\Big|_{r_+}^{r} \, , \label{aoah2}
\end{align}
\end{subequations}
where $a_O$ and $a_H$ are two constants, corresponding to the fluxes at
infinity and horizon, respectively. Similar to the case of the
gauge current, we express the total energy momentum tensor
as a sum of two combinations
$T_{~\nu}^{\mu} = T_{(O)\nu}^{\mu}\Theta(r) +T_{(H)\nu}^{\mu}H(r)$.
Using Eqs. (\ref{aoah1}) and (\ref{aoah2}), we find
\bea
 \sqrt{-g}\nabla_{\mu}T_{~t}^{\mu} &=& c_O\p_r\mathcal{A}_t
  +\p_r\Big[\big(\frac{1}{4\pi}\mathcal{A}_t^2 +N_{~t}^{r}\big)H\Big] \nn \\
 && +\Big[\sqrt{-g}\big(T_{(O)t}^{r} -T_{(H)t}^{r}\big)
 +\frac{1}{4\pi}\mathcal{A}_t^2
 +N_{~t}^{r}\Big]\delta(r -r_+ -\varepsilon) \, .
\eea
In the above equation, the first term is the classical effect of
the background U(1) gauge field for constant current flow. The second
term should be cancelled by the quantum effect of the classically
irrelevant ingoing modes. In order to guarantee the energy momentum
tensor is invariant under general coordinate transformations,
the third term must vanish at the horizon, which yields
\be
 a_O = a_H +\frac{1}{4\pi}\mathcal{A}_t^2(r_+)
 +\frac{1}{192\pi}F^{\prime 2}(r_+) \, ,
\ee
where we have used $N_{~t}^{r}(r_+) = -F^{\prime 2}(r_+)/(192\pi)$.
As what we have done to evaluate the gauge current at infinity,
to fix $a_O$ completely, we require to impose the boundary
condition that the covariant energy momentum tensor vanishes at
horizon, i.e., $a_H = 0$. We will see that such a boundary condition
is compatible with the Unruh vacuum in the next section. Therefore,
the total flow of energy momentum tensor is
\be
 a_O = \frac{1}{4\pi}\mathcal{A}_t^2(r_+) +\frac{\kappa^2}{48\pi} \, , \qquad
 \kappa = \frac{1}{2}F^{\prime}(r_+)
 = \frac{r_+V'(r_+)}{4\sqrt{U(r_+)}} \, , \label{EMTflux}
\ee
For the sake of comparing the total energy momentum flux (\ref{EMTflux})
with the Hawking one, we consider Hawking radiation with the Fermionic
Plank distribution $N_{e,m}(\omega)= 1/(e^{[\omega-e\hat{\Phi}_+ -m\Omega(r_+)
]/T_H}+1)$ in the background of the EMCS-G\"{o}del black hole, where
$T_H$ is the Hawking temperature (\ref{HTsurgrav}) via surface gravity
formula, $\hat{\Phi}_+ =\hat{B}(r_+)+\Omega(r_+) C(r_+)$ is the electric
chemical potential of the gauge field (\ref{ReGauge}) at the horizon
and $\Omega(r_+)$ is the angular velocity at the horizon. The Hawking
flux with this distribution is
\be
F_M=\int_0^\infty \frac{d\omega}{2\pi} \omega
[N_{e,m}(\omega)+N_{-e,-m}(\omega)]
 =\frac{1}{4\pi}\mathcal{A}_t^2(r_+)+\frac{\kappa^2}{48\pi}
 \, , \label{Hawkingflu}
\ee
which takes the same form as Eq. (\ref{EMTflux}). This implies that we
have reproduced the Hawking temperature (\ref{HTsurgrav}) via the
covariant anomaly cancellation method.

\section{Hawking fluxes and effective action}\label{hfeffa}

In this section, we will use the effective action method to exploit
Hawking radiation of the EMCS-G\"{o}del black hole in background of
the two dimensional metric (\ref{effmetr}) and gauge field
(\ref{effgau}). In two dimensional effective theory, there exist
normal effective action and chiral effective action. The former describes
the effective theory away from the event horizon. The energy momentum tensor
and gauge current induced from this action are anomaly free and take
consistent forms. The normal effective action has been used to derive
the Hawking fluxes of the Reissner-Nordstr\"{o}m black hole \cite{RotatingAnom}.
On the other hand, the chiral effective action \cite{BanerKeff,HLca}
depicts the chiral theory, in which the energy momentum tensor and gauge
current are not conserved but covariantly anomalous. By adopting the
covariant boundary condition at the event horizon, this effective action
can be applied to compute the Hawking fluxes of black holes \cite{BanerKeff}.
In our work \cite{Wueff}, the chiral effective action method has been
extended to reproduce the Hawking fluxes of the Schwarzschild black
holes in the isotropic coordinates where the determinant of the metric
vanishes at the horizon.

\subsection{Normal effective action and Hawking fluxes}\label{hfeffa1}

In two dimensional effective theory, the normal effective action is obtained by functional
integration of the conformal anomaly \cite{RotatingAnom,TwoDefBF}. It consists
of the gravitational (Polyakov) part and the gauge part. From a variation of
this effective action, we get the energy momentum tensor and gauge current
\cite{RotatingAnom,BanerKeff}
\bea
T_{\mu\nu}
 &=& -\frac{1}{\pi}\Big(\nabla_{\mu}\mathcal{B}\nabla_{\nu}\mathcal{B}
 -g_{\mu\nu}\frac{1}{2}\nabla^{\rho}\mathcal{B}\nabla_{\rho}\mathcal{B}\Big)  \nn \\
 &&-\frac{1}{48\pi}\Big[\nabla_{\mu}\mathcal{G}\nabla_{\nu}\mathcal{G}
 -2\nabla_{\mu}\nabla_{\nu}\mathcal{G}
  +g_{\mu\nu}\big(2R
  -\frac{1}{2}\nabla^{\rho}\mathcal{G}\nabla_{\rho}\mathcal{G}\big)\Big]
  \, , \label{NeffaEMT} \\
J^\mu &=& \frac{1}{\pi \sqrt{-g}}\epsilon^{\mu\nu}\partial_\nu\mathcal{B}
  \, , \label{NeffaCurr}
\eea
where $R=-F^{\prime\prime}(r)$ is the Ricci scalar of the metric
(\ref{effmetr}), and the two auxiliary fields $\mathcal{B}$ and $\mathcal{G}$
satisfy
\be
\nabla^{\mu}\nabla_{\mu}\mathcal{B}
 =-\frac{\epsilon^{\mu\nu}}{ 2\sqrt{-g}}\mathcal{F}_{\mu\nu} \, , \qquad
\nabla^{\mu}\nabla_{\mu}\mathcal{G}
=R \, . \label{auxilfields}
\ee
From Eqs. (\ref{NeffaEMT}) and (\ref{NeffaCurr}), we find that the gauge
current takes the conserved form $\nabla_{\mu}J^\mu =0$ while the energy
momentum tensor obeys the Lorentz force law (\ref{gat1}) and the trace
anomaly
\be
 \nabla_{\mu}T^{\nu}_{~\nu} = \mathcal{F}_{\mu\nu}J^{\mu}
 \, , \qquad \qquad
T^{\mu}_{~\mu} =-\frac{R}{24\pi} \, .
\ee

In the background of metric (\ref{effmetr}) and gauge field
(\ref{effgau}), solving Eq. (\ref{auxilfields}), we get
\bea
\partial_t \mathcal{G} &=& a \, , \qquad
\partial_r \mathcal{G} =\frac{b-2K}{F(r)} \, ,
\qquad K=\frac{1}{2}F^\prime(r) \, , \label{CurrSoauxf} \\
\partial_t \mathcal{B} &=& \alpha \, , \qquad
\partial_r \mathcal{B} =\frac{\beta +\mathcal{A}_t(r)}{F(r)} \, ,
\eea
where parameters $a$, $b$, $\alpha$, $\beta$ are constants. They can be
determined by proper boundary conditions. As in the previous section,
we still choose the boundary conditions that are compatible with the
Unruh vacuum. Such a choice requires us to express the energy momentum
tensor and gauge current in the Eddington-Finkelstein coordinate
system $\{u,v\}$, where $u=t-r_*$, $v=t+r_*$, and $dr_* =dr/F(r)$.
We have
\bea
T_{uu}&=& -\frac{1}{4\pi}(\alpha-\beta-\mathcal{A}_t(r))^2
 -\frac{1}{192\pi}\big[(a-b)^2-4K^2+4F(r)K^\prime \big] \, , \\
T_{uv}&=& T_{vu}=-\frac{1}{96\pi} F(r)F^{\prime\prime}(r) \, ,\\
T_{vv}&=& -\frac{1}{4\pi}(\alpha+\beta+\mathcal{A}_t(r))^2
 -\frac{1}{192\pi}\big[(a+b)^2-4K^2+4F(r)K^\prime \big] \, , \\
J_u &=& \frac{1}{2\pi}(\alpha-\beta-\mathcal{A}_t(r)) \, , \qquad
J_v = -\frac{1}{2\pi}(\alpha+\beta+\mathcal{A}_t(r)) \, .
\eea
Adopting the Unruh vacuum boundary conditions
\begin{subequations}
\begin{align}
J_u &= 0 \, , \qquad T_{uu} =0 \, , \qquad \qquad r=r_+
\, ,  \label{UVbound1} \\
J_v &= 0 \, , \qquad T_{vv} =0 \, , \qquad \qquad r\rightarrow +\infty
 \label{UVbound2} \, ,
\end{align}
\end{subequations}
the constants $a$, $b$, $\alpha$, $\beta$ can be solved as
\be
a =-b =\pm \kappa \, , \qquad \qquad
\alpha =-\beta =\frac{1}{2} \mathcal{A}_t(r_+) \, . \label{evalfouc}
\ee

Substituting the four constants into the $(r,t)$-component of the energy
momentum tensor and the $r$-component of the gauge current, which
correspond to fluxes for Hawking radiation and the gauge field,
respectively, we obtain
\be
T^{r}_{~t} =\frac{1}{4\pi} \mathcal{A}_t(r_+)[ \mathcal{A}_t(r_+)
 -2 \mathcal{A}_t(r)] +\frac{\kappa^2}{48\pi} \, , \qquad
 J^r =-\frac{1}{2\pi} \mathcal{A}^2_t(r_+) \, ,
\ee
where $J^r$ is a constant since the normal effective action describes the
theory away from the horizon and the gauge current is conserved. Taking the
limit at infinity, we derive the charge flow and the Hawking fluxes
\bea
 J^r(r\rightarrow \infty) &=& -\frac{1}{2\pi} \mathcal{A}^2_t(r_+) \, ,
 \\
T^{r}_{~t}(r\rightarrow \infty) &=& \frac{1}{4\pi} \mathcal{A}^2_t(r_+)
 +\frac{\kappa^2}{48\pi} \, ,
\eea
which agree with Eqs. (\ref{gcurrca}) and (\ref{EMTflux}) via the covariant
anomalies in the previous section.

\subsection{Chiral effective action and Hawking fluxes}\label{hfeffa2}

Varying the chiral effective action, the covariant energy momentum
tensor $\widetilde{T}^{\mu}_{~\nu}$ and the covariant gauge current
$\widetilde{J}^\mu$ read \cite{BanerKeff,HLca}
\bea
\widetilde{T}_{\mu\nu}
 &=& -\frac{1}{4\pi}D_{\mu}\mathcal{B}D_{\nu}\mathcal{B}
 -\frac{1}{96\pi}\Big(\frac{1}{2}D_{\mu}\mathcal{G}D_{\nu}\mathcal{G}
 -D_{\mu}D_{\nu}\mathcal{G}
  +g_{\mu\nu} R \Big) \, , \\
\widetilde{J}^\mu &=& \frac{1}{2\pi }D^\mu\mathcal{B} \, ,
\eea
where the chiral covariant derivative $D_\mu =\sqrt{-g}\epsilon_{\mu\nu}D^\nu
=\nabla_\mu +\sqrt{-g}\epsilon_{\mu\nu}\nabla^\nu$,
$\widetilde{J}_\mu =\sqrt{-g}\epsilon_{\mu\nu}\widetilde{J}^\nu$,
and the two auxiliary fields $\mathcal{B}$ and $\mathcal{G}$ have been
defined by Eq. (\ref{auxilfields}). In the chiral effective theory, the
covariant energy momentum tensor and gauge current satisfy the anomalous
Ward identities,
\bea
 \nabla_\mu \widetilde{J}^{\mu}
 &=& \frac{-1}{4\pi\sqrt{-g}}\epsilon^{\rho \sigma}
 \mathcal{F}_{\rho \sigma} \, , \\
\nabla_{\mu}\widetilde{T}^{\mu}_{~\nu}
&=& \mathcal{F}_{\mu\nu}\widetilde{J}^{\mu}
 +\frac{1}{96\pi\sqrt{-g}}\epsilon_{\nu\mu}\p^{\mu}R \, .
\eea
The energy momentum tensor also obeys the covariant trace anomaly
$\widetilde{T}^{\mu}_{~\mu} =-R/(48\pi)$. Operating the chiral
covariant derivative on the auxiliary fields $\mathcal{G}$ and
$\mathcal{B}$, we get
\bea
D_t \mathcal{G} &=& -F(r)D_r \mathcal{G}
=\tilde{a}-\tilde{b}+2K  \, ,  \\
D_t \mathcal{B} &=& -F(r)D_r \mathcal{B}
= \tilde{\alpha}-\tilde{\beta}-\mathcal{A}_t(r) \, ,
\eea
where $\tilde{a}$, $\tilde{b}$, $\tilde{\alpha}$ and $\tilde{\beta}$
are constants. Their relations will be determined later. Now the
$(r,t)$-component of the covariant energy momentum tensor and the
covariant gauge current can be read off as
\bea
\widetilde{T}^r_{~t}&=& \frac{1}{4\pi}(\tilde{\alpha}-\tilde{\beta}
 -\mathcal{A}_t(r))^2
 +\frac{1}{192\pi}\big[(\tilde{a}-\tilde{b})^2-4K^2+4F(r)K^\prime \big] \, , \\
\widetilde{J}^r &=& F(r)\widetilde{J}^t
=-\frac{1}{2\pi}(\tilde{\alpha}-\tilde{\beta}
-\mathcal{A}_t(r)) \, .
\eea
Here we do not present the other components of the energy momentum tensor
$\widetilde{T}^{\mu}_{~\nu}$, which are useless for computation of the Hawking flux.
Finally, to derive the Hawking fluxes and currents for gauge fields, we need to
impose the covariant boundary conditions that the covariant energy
momentum tensor and gauge current vanish at the horizon \cite{BanerKeff},
namely,
\be
\tilde{\alpha}=\tilde{\beta}+\mathcal{A}_t(r_+) \, , \qquad \qquad
\tilde{a}=\tilde{b}\pm 2\kappa \, . \label{covboun}
\ee
Therefore, taking the asymptotic limit, we obtain the gauge currents and
the fluxes for energy momentum tensor
\bea
 \widetilde{J}^r(r\rightarrow \infty) &=& -\frac{1}{2\pi} \mathcal{A}^2_t(r_+) \, ,
 \\
\widetilde{T}^{r}_{~t}(r\rightarrow \infty) &=& \frac{1}{4\pi} \mathcal{A}^2_t(r_+)
 +\frac{\kappa^2}{48\pi} \, .
\eea
They are in agreement with Eqs. (\ref{gcurrca}) and (\ref{EMTflux}).

It is worth noting that the covariant boundary conditions adopted in this subsection
and the previous section are compatible with
the Unruh vacuum. To see this, as in the case of the normal effective action,
we could express the energy momentum tensor and gauge current in the
Eddington-Finkelstein coordinate system $\{u,v\}$.
By changing
$(\alpha,\beta,a,b)$ to $(\tilde{\alpha},\tilde{\beta},\tilde{a},\tilde{b})$,
respectively,
we find $\widetilde{T}_{uu} =T_{uu}$,
$\widetilde{T}_{uv} =T_{uv}/2$, $\widetilde{T}_{vv} =0$, $\widetilde{J}_{u}
=J_{u}$, and $\widetilde{J}_{v}=0$, where $\widetilde{T}_{vv}$ and
$\widetilde{J}_{v}$ are zeros, since the two dimensional effective theory
is chiral and there are no ingoing modes. Clearly, Eqs. (\ref{UVbound1})
and (\ref{UVbound2}) hold when one adopts the covariant boundary conditions
(\ref{covboun}).

\section{Summary}

We have obtained the EMCS-G\"{o}del black hole's \cite{SQWu} conserved charges,
such as the mass, the angular momenta and the electric charge
along the lines of \cite{TherBC}. They
are consistent with the differential first law and the generalized
integral Smarr formula of black hole thermodynamics provided that
the G\"{o}del parameter $j$ is a thermodynamical variable.
The EMCS-G\"{o}del black hole is an exact charged rotating solution
in the five dimensional minimal supergravity with G\"{o}del background.
It has closed time-like curves through every point and the asymptotic
structure with a global rotation. These peculiar properties lead to
the failure of the traditional methods on calculating the conserved
charges. A viable method was presented in \cite{TherBC}. Whether there
exist other methods to compute the conserved charges of the
EMCS-G\"{o}del black hole is still open.

Besides, we have derived the Hawking fluxes of the EMCS-G\"{o}del black hole
via covariant gravitational and gauge anomalies, as
well as the effective action. By applying the technique of dimensional
reduction to the metric (\ref{GBHansatz}) and the gauge field (\ref{ReGauge}),
which has the vanishing electro-static potential at infinity,
higher dimensional theory near the horizon can be effectively described
by a two dimensional theory in the background of metric (\ref{effmetr})
and gauge field (\ref{effgau}). The reduced gauge field consists of
two U(1) fields, one from the original gauge field and the other
from the axial isometry along the $\phi$ direction. On the basis of both
the two dimensional metric and gauge field, by adopting the covariant
boundary conditions that are compatible with the Unruh vacuum,
the covariant anomaly cancellation method and the approach of the effective
action, including the normal and chiral effective action, were used
to derive the same Hawking fluxes as those from Planck distribution for
blackbody radiation in the background of the EMCS-G\"{o}del black hole.

Our results show that Hawking radiation is a quantum phenomenon taking
place at the event horizon, since both the methods of the anomaly
cancellation and the chiral effective action only rely on the quantum
anomalies and boundary conditions at the event horizon. Besides, our
calculation supports the anomaly cancellation method is applicable to
the black holes in the five dimensional minimal supergravity with
G\"{o}del background. In some sense, the approach of the anomaly
cancellation is universal except for the procedure of dimensional
reduction for different background spacetime considered in each case.
A further development of this work is to derive the entropy of the
EMCS-G\"{o}del black hole in the same two dimensional effective theory,
like in \cite{LeoRTuna,ViroBGK}. Our analysis in the present paper can
be directly generalized to the squashed charged rotating black hole in the five
dimensional G\"{o}del universe \cite{SquaGBH}.

\medskip
\textbf{Acknowledgments}:
S.Q. Wu thanks Dr. G. Comp\`{e}re for helpful discussions, particularly for sharing his
program to calculate conserved charges. This work was partially supported by the Natural
Science Foundation of China under Grant Nos. 10975058 and 10675051. J.J. Peng was also
supported in part by a Graduate Innovation Foundation of HuaZhong Normal University.


\end{document}